\documentclass[12pt]{article}
\usepackage{graphicx}
\usepackage{xcolor}
\usepackage[hidelinks]{hyperref}
\hypersetup{
    colorlinks,
    linkcolor={red!50!black},
    citecolor={blue!50!black},
    urlcolor={blue!80!black}
}
\usepackage{amsmath}
\usepackage{cleveref}

\newcommand\pubdate{\today}

\newcommand{\orcid}[1]{\,\href{https://orcid.org/#1}{\includegraphics[width=9pt]{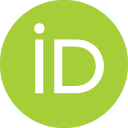}}}
\newcommand{\orcidTK}{0000-0002-7516-8292} %
\newcommand{\orcidTJ}{0000-0002-1334-7607} %
\newcommand{\orcidPD}{0000-0001-7960-7953} %
\newcommand{\orcidMK}{0000-0002-4665-3088} %
\newcommand{\orcidKK}{0000-0003-1412-447X} %
\newcommand{\orcidAK}{0000-0002-4090-0084} %
\newcommand{\orcidJM}{0000-0001-9343-9351} %
\newcommand{\orcidFO}{0000-0001-6799-2436} %
\newcommand{\orcidIS}{0000-0003-0373-474X} %
\newcommand{\orcidJY}{0000-0001-8366-0968} %
\newcommand{\orcidRR}{0000-0002-3316-2175} %
\newcommand{\orcidPR}{0000-0002-8570-5506} %
\newcommand{\orcidCL}{0000-0001-7509-5655} %
\newcommand{\orcidND}{0000-0003-0962-631X} %
\def\smu{{Department of Physics, Southern Methodist University,
    Dallas, TX 75275-0175, U.S.A.}}
\def\jlab{{Jefferson Lab, Newport News, VA 23606, U.S.A.}}
\def\krakow{{Institute of Nuclear Physics Polish Academy of Sciences, PL-31342 Krakow, Poland}}
\def\muenster{{Institut f{\"u}r Theoretische Physik, Westf{\"a}lische Wilhelms-Universit{\"a}t M{\"u}nster,
  \\Wilhelm-Klemm-Stra{\ss}e 9, D-48149 M{\"u}nster, Germany}}
\def\lpsc{{Laboratoire de Physique Subatomique et de Cosmologie, Université Grenoble-Alpes, 
    \\CNRS/IN2P3, 53 avenue des Martyrs, 38026 Grenoble, France}}
\def\fnal{{Fermi National Accelerator Laboratory, Batavia, Illinois 60510, USA}}
\def\jyv{{University of Jyväskylä, Department of Physics, P.O.\ Box 35, FI-40014 University of Jyväskylä, Finland}}
\def\helsinki{{Helsinki Institute of Physics, P.O.\ Box 64, FI-00014 University of Helsinki, Finland}}

\def\Title#1{\begin{center} {\Large #1 } \end{center}}
\def\Author#1{\begin{center}{ \sc #1} \end{center}}
\def\Address#1{\begin{center}{ \it #1} \end{center}}

\newcommand\pubblock{\rightline{\begin{tabular}{l}  \\ 
         \pubdate  \end{tabular}}}
\newenvironment{Abstract}{\begin{quotation}  }{\end{quotation}}
\newenvironment{Presented}{\begin{quotation} \begin{center} 
             PRESENTED AT\end{center}\bigskip 
      \begin{center}\begin{large}}{\end{large}\end{center} \end{quotation}}

\begin{document}
\begin{titlepage}
 \pubblock
\vfill
\Title{Towards a New nCTEQ global nPDF release}
\vfill
\Author{P. Risse$^{a,\dag}$\orcid{\orcidPR}, N. Derakhshanian$^e$\orcid{\orcidND}, P.~Duwent\"aster$^{b,c}$\orcid{\orcidPD}, T.~Je\v{z}o$^a$\orcid{\orcidTJ}, C.~Keppel$^d$\orcid{\orcidTK}, M.~Klasen$^a$\orcid{\orcidMK}, K.~Kova\v{r}\'{i}k$^a$\orcid{\orcidKK}, A.~Kusina$^e$\orcid{\orcidAK}, C.~L\'eger$^f$\orcid{\orcidCL}, J.G.~Morf\'{i}n$^g$\orcid{\orcidJM}, F.I.~Olness$^h$\orcid{\orcidFO}, R.~Ruiz$^e$\orcid{\orcidRR}, I.~Schienbein$^f$\orcid{\orcidIS}, J.Y.~Yu$^f$\orcid{\orcidJY}}

\Address{\scriptsize $^a$\muenster \\ $^b$\jyv \\ $^c$\helsinki \\ $^d$\jlab \\ $^e$\krakow \\ $^f$\lpsc \\ $^g$\fnal\\$^h$\smu} 
\vfill
\begin{Abstract}
We discuss the foundation for a new  global nCTEQ nuclear PDF analysis, combining a number of our previous analyses into one consistent framework with updates to the underlying theoretical treatment as well as the addition of new available data. In particular, the new global release will be the first nCTEQ release containing neutrino DIS scattering data in a consistent manner together with JLab high-$x$ DIS data and new LHC p-Pb data. These additions will improve the data-driven description of nuclear PDFs in new regions, especially the strange quark and the gluon PDF at low-$x$.
\end{Abstract}
\vfill
\let\thefootnote\relax\footnotetext{$^\dag$\texttt{risse.p@uni-muenster.de}}
\begin{Presented}
DIS2023: XXX International Workshop on Deep-Inelastic Scattering and
Related Subjects, \\
Michigan State University, USA, 27-31 March 2023 \\
     \includegraphics[width=9cm]{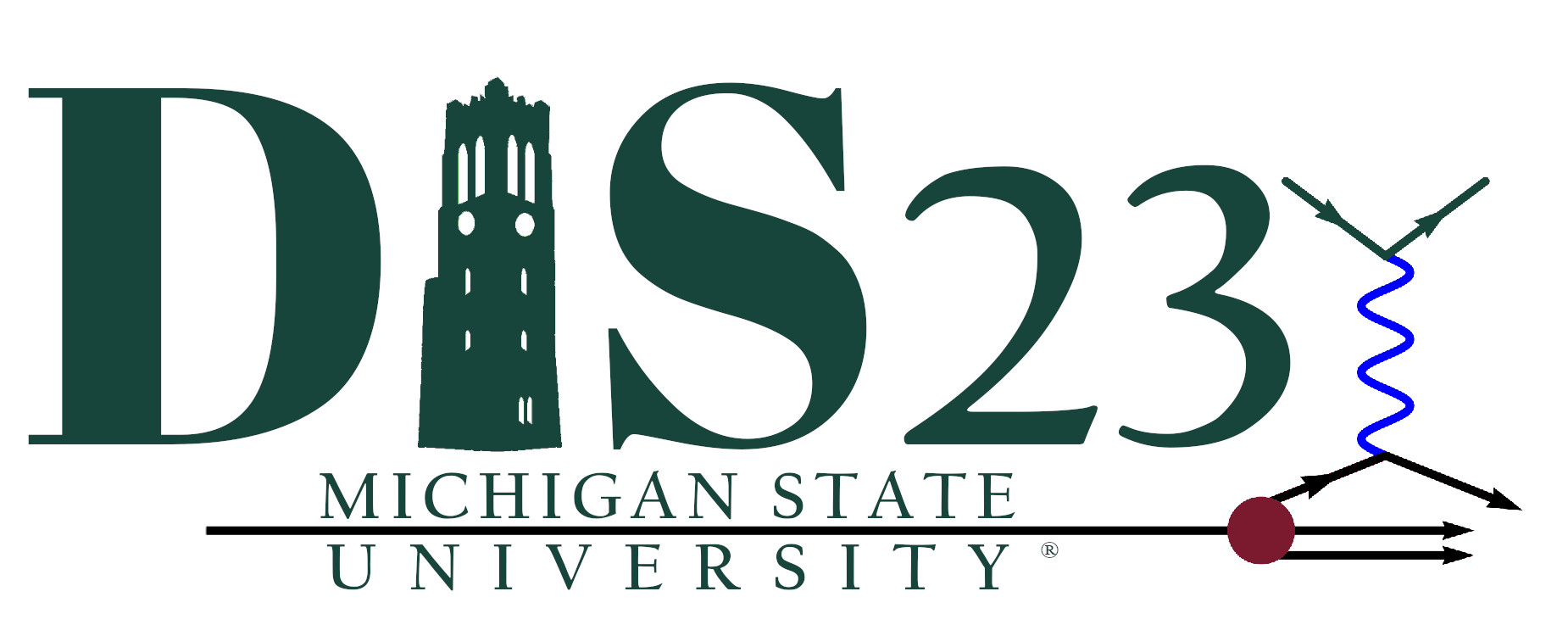}
\end{Presented}
\vfill
\end{titlepage}

\section{Motivation}
Parton Distribution Functions (PDFs) describe the structure of hadrons as composed of quarks and gluons. They are one of the key elements in making predictions for collider experiments in the theoretical framework of perturbative quantum chromodynamics (QCD). So far, PDFs cannot be reliably computed from first principles. Only the evolution with the energy scale $Q$ is calculable through the DGLAP evolution equations. Thanks to the QCD factorization theorem, it is possible to use their universality and determine the $x$-dependence at some initial scale $Q_0$ by comparing PDF-dependent predictions with experimental measurements. This is a process called `global QCD analysis' \cite{Kovarik:2015cma}. In the nuclear case additional effects such as shadowing, antishadowing, the EMC effect and Fermi motion need to be parameterized as well. 

Several collaborations perform global analyses of nuclear PDFs (e.g. \cite{AbdulKhalek:2022fyi,Duwentaster:2022kpv,Eskola:2021nhw}).
In the nCTEQ framework the parametrisation of nuclear PDFs is done in two steps. First, at the initial scale $Q_0=1.3$ GeV, the so-called bound proton PDF, $f^{p/A}$, is parametrized as
\begin{align}
    \label{eq:PDF_param}
    xf^{p/A}_i(x,Q_0) \,\,&= \,\,c_0x^{c_1}(1-x)^{c_2}e^{c_3 x}(1+e^{c_4}x)^{c_5},\\
    \label{eq:nuclear_param}
    c_k \,\,&\rightarrow \,\,p_k + a_k\left(1-A^{-b_k}\right)
\end{align}
and depends explicitly on the nuclear mass number $A$. The bound neutron PDFs $f^{n/A}$ are obtained by making use of isospin symmetry and the full nucleus is constructed via
\begin{equation}
    f^{(A,Z)}_i(x,Q) = \frac{Z}{A}f^{p/A}_i(x,Q) + \frac{A-Z}{A}f^{n/A}_i(x,Q).
\end{equation}
The nCTEQ15 fit, the latest global release from the nCTEQ collaboration, used data from deep-inelastic scattering (DIS) and Drell-Yan (DY) to constrain the quark and anti-quark nPDFs as well as data from RHIC neutral pion production for the gluon. From today's perspective the fit had a number of limitations including the precision in the high-$x$ region, the gluon and  the strange PDF. Lowering the energy cuts in the data opens up new constraints on the high-$x$ region for the up and down quarks. LHC data and a reanalysis of neutrino DIS and charm dimuon data allows us to improve this fit in the gluon and strange directions. We design a new global fit built upon these results.

\section{New proton Baseline and nuclear parametrization}

The new release will use an improved proton baseline from the ``CTEQ-Jefferson Lab" (CJ) proton release \cite{Accardi:2016qay}. This will update the current CTEQ6 baseline. The parametrization 
\begin{equation}
    xf^{p/A}_i(x,Q_0) \,\,= \,\, c_0x^{c_1}(1-x)^{c_2}(1+c_3\sqrt{x}+c_4x)
\end{equation}
itself is simpler than \cref{eq:PDF_param}, leading to less correlated parameters in the fit. Furthermore, CJ made use of constraints from newer data and includes a careful treatment of deuterium corrections and high-$x$ effects, making it especially suited for an extension to the nuclear dimension. 

The second parametric update is a new $A$-dependence of the parameters: 
\begin{equation}
    c_k \,\,\rightarrow \,\,p_k + a_k\ln(A) +b_k\ln^2(A).
\end{equation}
Compared to \cref{eq:nuclear_param} the correlations are limited between the parameters yielding a more accessible $\chi^2$-function and a more truthful representation of uncertainties. 

\section{Relaxed kinematical cuts}

\begin{figure}
    \centering
    \includegraphics[width=0.7\textwidth]{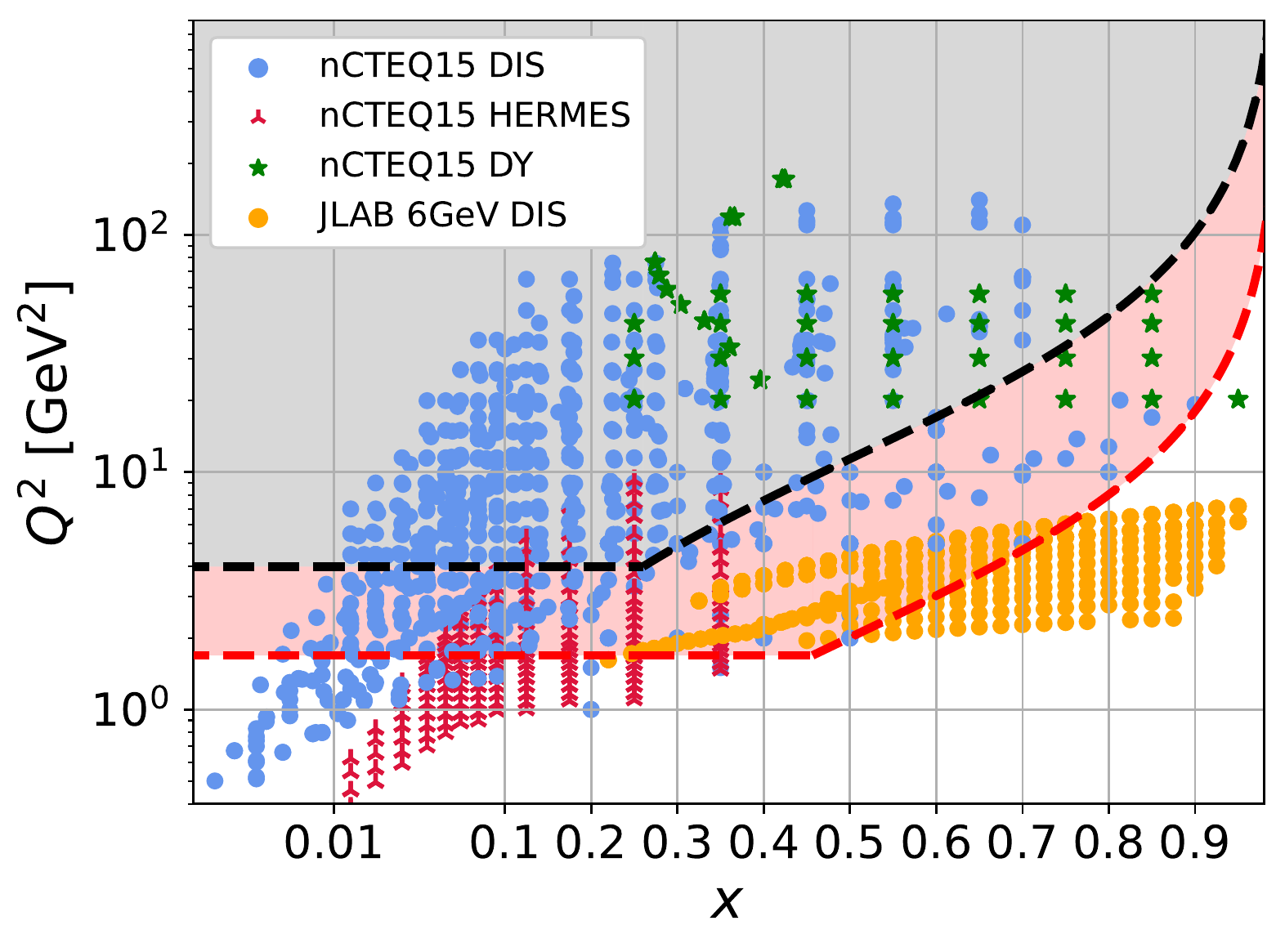}
    \caption{Coverage of the kinematic $(x,Q^2)$-plane of the experimental data sets. The nCTEQ15 $Q$- and $W$-cut (black dashed line) removed most of the HERMES DIS data and all of the data from JLAB. The improved cut is shown as the red dashed line.}
    \label{fig:hix_data-cuts}
\end{figure}
The nCTEQ15 fit used restrictive cuts for the DIS data with $Q = 2$ GeV and $W = 3.5$ GeV. Whilst this allows for a simplified theoretical prediction, it removes a significant amount of data points: a large fraction of the data from HERMES and all of the JLAB data. The cut is depicted by the black dashed line in fig.\,\ref{fig:hix_data-cuts}. Relaxed cuts within the nCTEQ15HIX analysis \cite{Segarra:2020gtj}, shown as the red dashed line, allowed us to include these data and therefore improve the constraints on the high-$x$ region. In total, about 850 new data points were added. The most notable improvement is the reduction in uncertainties in the up- and down-sector for $x$-values between $10^{-2}$ and 0.7. 

The new cuts of $Q=1.3$ GeV and $W=1.7$ GeV are possible due to a careful treatment of target mass corrections, the inclusion of higher twist effects and an improved description of Deuterium. The details of the target mass corrections have been worked out for different nuclei in Ref.\,\cite{Ruiz:2023ozv} and lead to corrections of up to 15\% in the $F_2$ structure function at $Q=1.3$ GeV. At this energy the span between different nuclei is at about 2\% from the lightest to the heaviest nuclei. The effects decrease quickly with higher energies.

A parametrization of the higher twist effects has been worked out in the CJ proton release. It leads to a reduction in the intermediate-$x$ region ($x\sim 0.3$) and becomes large and positive at large $x$. A possible nuclear dependence of these corrections has been tested but had very limited impact.

A large fraction of the nuclear DIS data are expressed as a $F^A_2/F^D_2$ ratio of structure functions on a heavy nuclear target and on a deuteron target. The deuteron structure is more complex than a isoscalar combination of proton and neutron PDFs. Although a sizable part of the deuteron system can be captured by these charge factors weighting the quark PDFs, the residual dynamical nuclear effects are not negligible. Particularly at large-$x$ values deviations of up to 7\% exist.

\section{Heavy-quark and quarkonium data}
The experiments at the LHC have provided data on $W/Z$ boson and single inclusive hadron production in $pA$ scattering. This improved the determination of the gluon PDF within the nCTEQ15WZ \cite{Kusina:2020lyz} and nCTEQ15WZSIH \cite{Duwentaster:2021ioo} global fits. The uncertainty was reduced in both fits by a factor of two in the low-$x$ region: $x\sim10^{-3}-10^{-2}$. Also, the central value of the strange PDF was affected but the uncertainty remained substantial.

Heavy quarkonium and open heavy-flavor meson production data can constrain the nuclear gluon further. The gluon contribution dominates the overall cross section of these processes and a naive leading order estimation of
\begin{equation}
    x \approx \frac{2p_T}{\sqrt{s}}\exp(-|y|)
\end{equation}
shows that the data is sensitive to $x$ values below $10^{-5}$. A fast evaluation of the cross section is possible by the data driven approach of the Crystal Ball function \cite{Kusina:2017gkz,Lansberg:2016deg}. The ansatz assumes that the cross section is dominated by the gluon and its parameters are inferred via the well measured proton-proton cross sections. The prediction has been verified by the next-to-leading order calculations in QCD \cite{Kniehl:2012ti} and NRQCD \cite{Butenschoen:2010rq}.
\begin{figure}
    \centering
    \includegraphics[width=0.9\textwidth]{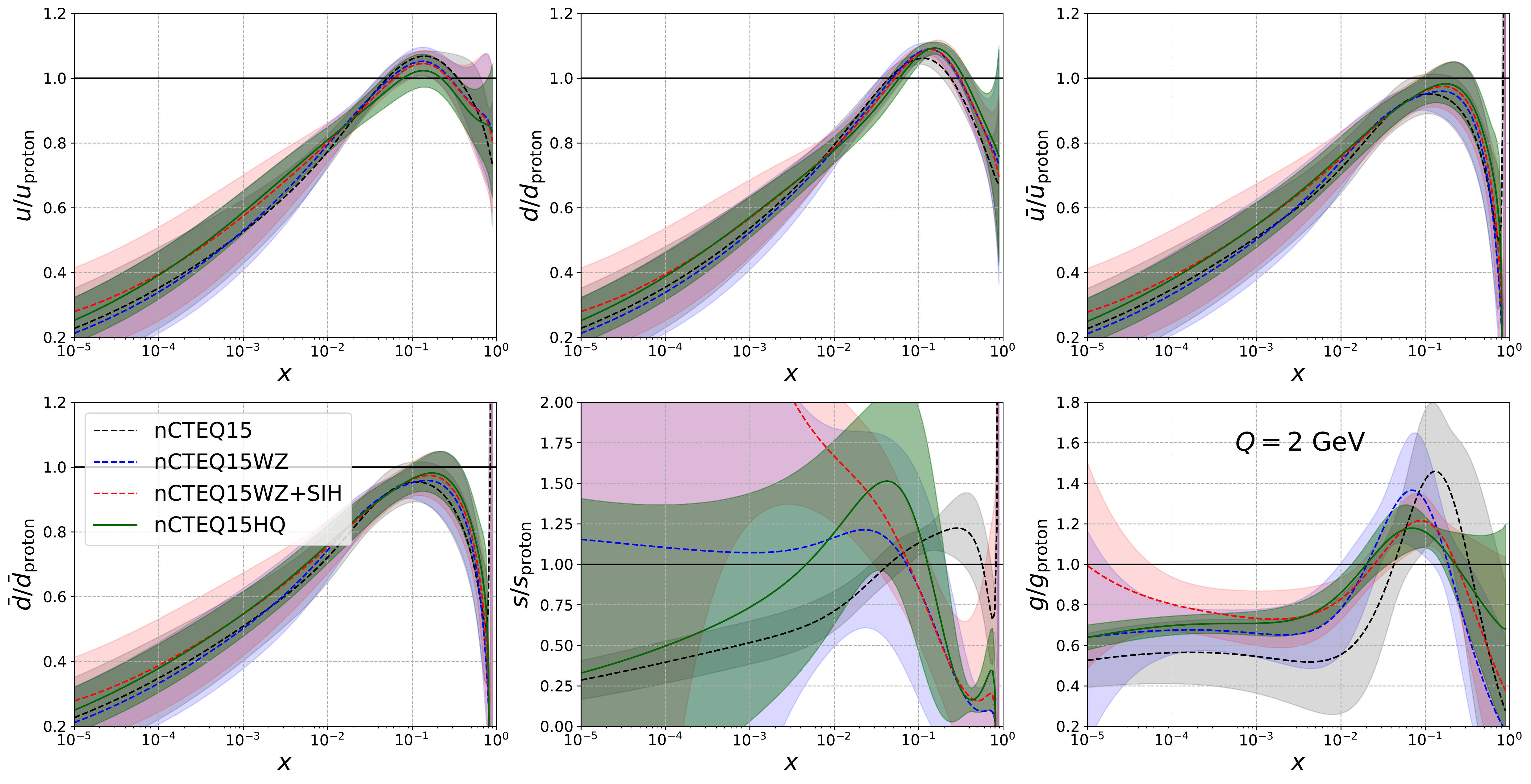}
    \caption{Ratio lead to proton PDFs from the nCTEQ15 fit (black), the nCTEQ15WZ fit (blue) and the nCTEQ15WZSIH (red). The new nCTEQ15HQ fit is shown in green.}
    \label{fig:nCTEQ15HQ_ratio_lead}
\end{figure}

In \cref{fig:nCTEQ15HQ_ratio_lead} the resulting nCTEQ15HQ fit, shown in green, is compared to the nCTEQ15 in black, the nCTEQ15WZ in blue and the nCTEQ15WZSIH in red.  Clearly, the central value in the up- and down-sector is not affected whereas the uncertainty is reduced. The gluon is impacted the most. Its fit results in a shifted central value compatible with the former results whilst yielding a significantly reduced uncertainty.

\section{Neutrino DIS and dimuon data}
As the uncertainty of the strange PDF remains large, other additional data needs to be utilized. Charged Current neutrino DIS and charm dimuon production data have long been known to be sensitive to the strange PDF. However, the compatibility of the corresponding measurements from CDHSW, CCFR, NuTeV and Chorus and also with neutral current DIS measurements exhibits tensions. 

In a global fit in Ref.\,\cite{Muzakka:2022wey} an extensive compatibility study was carried out both within the charged current data sets and those used in nCTEQ15WZSIH. Even with an improved calculation of the deuteron structure and a larger $\Delta\chi^2$ tolerance of 45 instead of 35 the tension persists. One way of removing the tension is to use a kinematical cut of $x>0.1$ but a better result was obtained by using only Chorus and dimuon data. 

Overall the central value of the PDFs from the best neutrino fit (BaseDimuChorus)  is only affected in the case of the strange quarks whilst all uncertainties are decreased. At the same time the data has an impact on the uncertainty of the valence PDFs but they are dominantly constrained by the neural current DIS experiments.  

\section{Conclusion}
In conclusion, the next nCTEQ global fit will extend the nCTEQ15 results in several directions. 
An update in the proton baseline, i.e. the CJ PDFs, will provide improved proton boundary conditions. The parametrization update on the level of nuclear modifications (obtained by the new $A$-dependence) provides a better disentanglement between individual parameters which simplifies the fitting procedure and results in a more truthful representation of PDF uncertainties.

Relaxed kinematical cuts in the DIS data with a careful treatment of target mass corrections, higher twist effects and improved prediction for the deuterium kinematics yield stronger constrains in the high-$x$ region throughout all flavors. Additionally it opens up the possibility of including new data from HERMES and JLAB increasing the coverage of the data sets overall.

The LHC has produced precise data on $W/Z$, heavy quark and quarkonium production in $pA$ collisions. We will include the new data sets, where the heavy quark and quarkonium production provides access to very low values of $x$ in heavy nuclei and is accurately described by a data-driven approach. In consequence the new fit will include refined constraints on the gluon PDF even down to $10^{-5}$ in the momentum fraction.

Finally, improved flavor separation is gained by incorporating fixed-target neutrino DIS and charm dimuon production data with high statistics. By concentrating only on the CHORUS and dimuon data the compatibility with neutral current DIS is obtained. It yields better constraints on the strange quark and also reduces uncertainties on the up- and down-quarks.

\footnotesize
\bibliographystyle{utphys}
\bibliography{references}
\end{document}